\documentclass[sigconf]{acmart}

\acmConference[ICSE 2024]{46th International Conference on Software Engineering}{April 2024}{Lisbon, Portugal}

\usepackage{listings}
\usepackage{color}
\usepackage{makecell}
\usepackage{float}

\usepackage{eso-pic}
\usepackage{xcolor}

\newcommand\AtPageUpperMyright[1]{%
  \AtPageUpperLeft{%
    \put(\LenToUnit{0.09\paperwidth},\LenToUnit{-1cm}){%
      \parbox{1.0\textwidth}{\centering\textbf{\textcolor{red}{#1}}}%
    }%
  }%
}
\AddToShipoutPictureBG{%
  \AtPageUpperMyright{Accepted for publication at the LLM4Code workshop in the 46th International Conference on Software Engineering (ICSE 2024)}
}

\definecolor{dkgreen}{rgb}{0,0.6,0}
\definecolor{gray}{rgb}{0.5,0.5,0.5}
\definecolor{mauve}{rgb}{0.58,0,0.82}

\AtBeginDocument{%
  }

\setcopyright{acmlicensed}
\copyrightyear{2023}
\acmYear{2024}
\acmDOI{XXXXXXX.XXXXXXX}

\acmConference[ICSE]{ International Conference on Software Engineering 2024, Workshop LLM4Code}{2024}{Lisbon, Portugal}
\acmISBN{978-1-4503-XXXX-X/18/06}

\begin{document}




\title{An Empirical Study on Usage and Perceptions of LLMs in a Software Engineering Project}

\author{Sanka Rasnayaka, Guanlin Wang, Ridwan Shariffdeen, Ganesh Neelakanta Iyer}
\affiliation{%
  \institution{School of Computing, National University of Singapore}
  \city{\{sanka@nus.edu.sg, wguanlin@u.nus.edu, ridwan@comp.nus.edu.sg, gni@nus.edu.sg\}}
  \country{}
}




%
\begin{abstract}
Large Language Models (LLMs) represent a leap in artificial intelligence, excelling in tasks using human language(s). Although the main focus of general-purpose LLMs is not code generation, they have shown promising results in the domain. However, the usefulness of LLMs in an academic software engineering project has not been fully explored yet. In this study, we explore the usefulness of LLMs for 214 students working in teams consisting of up to six members. Notably, in the academic course through which this study is conducted, students were encouraged to integrate LLMs into their development tool-chain, in contrast to most other academic courses that explicitly prohibit the use of LLMs.

In this paper, we analyze the AI-generated code, prompts used for code generation, and the human intervention levels to integrate the code into the code base. We also conduct a perception study to gain insights into the perceived usefulness, influencing factors, and future outlook of LLM from a computer science student's perspective. Our findings suggest that LLMs can play a crucial role in the early stages of software development, especially in generating foundational code structures, and helping with syntax and error debugging. These insights provide us with a framework on how to effectively utilize LLMs as a tool to enhance the productivity of software engineering students, and highlight the necessity of shifting the educational focus toward preparing students for successful human-AI collaboration.
\end{abstract}





\maketitle

\section{Introduction}
Large Language Models (LLMs) are a recent advancement in artificial intelligence \cite{Hadi2023}, \cite{Raiaan2023} designed to understand, generate, and manipulate human languages. Applications of LLMs range from simple tasks like text auto-completion to complex operations such as language translation. Even though the general purpose LLMs are not specifically trained for code generation tasks, they have shown impressive results in code generation, bug detection, and automated documentation \cite{LLMSE1}, \cite{10109345}. The possibility of translating natural language descriptions into executable code is an exciting prospect, since this can enable any layperson to become a programmer by bridging the gap between conceptual design and technical implementation. However, the usefulness of LLMs in a software engineering project is not well explored in the literature. 

In this paper, we explore the possible uses of LLMs for software development projects in a controlled academic setting. For this purpose, we use a software engineering course with an enrolment of more than 200 students at the School of Computing, National University of Singapore (NUS), to develop a C++ application with a team consisting of at most six students. In particular, we encourage students to leverage the benefits of LLMs in their development workflows in order to study the usefulness of LLMs within a software development life cycle. 


Throughout the software development process, we capture AI-generated code, the prompts used for code generation, and also record details about the level of human intervention performed on the AI-generated code. These help us to identify the additional effort required to integrate the AI-generated code into the project, as well as how to incorporate the support provided by the LLMs. These artifacts are automatically extracted from students' code repositories. In addition, we ask the students to provide us with insights into the usage patterns of LLMs for the development of their software engineering project. At the end of the project, we surveyed the students to gather perceptions of the usage of LLMs and expectations of an LLM for code generation. The survey is designed based on the well-established Unified Theory of Acceptance and Use of Technology (UTAUT) model \cite{V2003} for a systematic analysis of users using the code generators.

Analysing the software artifacts reveals that students found LLM code generation very helpful for getting started on the project, specifically for creating basic design patterns, data-structures and algorithms, getting C++ specific syntax help for simple constructs, and receiving feedback for their own code. These are captured in the prompt they used during the project development, where 22.95\% of student-written prompts were used for simple tasks and getting help for syntax issues in C++, while 39.34\% of prompts were used to generate solutions for common algorithms such as depth-first-search (DFS). Students also used LLMs as a pair-programmer, to easily understand and debug error messages they encounter, to improve their code quality and to verify adherence to best practices. 

The perception analysis shows that students found LLMs useful in routine tasks and using them improved their productivity. However, there were also some concerns regarding how using LLMs might deprive them of gaining coding skills and how it might impact the software engineering job market.  All institutional guidelines were followed, and we received approval from the Departmental Ethics Review Committee (DERC) for our study before it was conducted.

Our findings give a blueprint on what are the most useful use cases of AI-generated code in an academic software engineering project. The code analysis across project phases and categorization of AI-generated code shows how LLMs could be utilized as a tool that aids in the software development process, providing more capabilities to developers and improving the productivity of software engineers. This paper makes the following contributions:
\begin{itemize}
    \item To the best of our knowledge we present the first study on using LLMs for a team-based software engineering project and provide insights on how useful LLMs are in an academic setting. 
    \item We analyze the perceptions of students in using LLMs for software development and identify valuable insights that can be helpful to further LLM usage in teaching environments. 
    \item We provide all artifacts including AI-generated code, prompts used by the students, and survey results for further analysis by the research community.\footnote{https://github.com/WangGLJoseph/LLM-SE-Project-Study} 

\end{itemize}

The next section details the settings of the software engineering project, details of the artifact extraction, description of the user study, and results of the analysis. 

\section{Methodology}\label{sec-case}

\subsection{Project Description}


The CS3203 software engineering course conducted at NUS involves a semester-long team project. Each team is required to develop a Static Program Analyser (SPA) for programs developed using a custom programming language: \textit{SIMPLE}\footnote{https://nus-cs3203.github.io/course-website/contents/basic-spa-requirements/simple-programming.html}.  We designed this project to ensure students will experiment with SPA techniques and experience a realistic software development project. The scope of the project is carefully chosen such that the overall workload is manageable for a team of six students to complete within one semester, which is roughly 13 weeks. The students enrolled in this course were full-time undergraduate students between the ages of 20 to 24.

\textit{SIMPLE} is not a programming language to solve real-world programming challenges, but it contains the basic constructs of a programming language. It can be used to write meaningful programs and is syntactically very similar to Python language in its control constructs, such as \verb+if+ and \verb+while+. 

SPA is an interactive program that automatically answers queries about a provided \textit{SIMPLE} program by:
\begin{itemize}
    \item Extract details of the program written in \textit{SIMPLE}, by analysing design entities and other abstractions. 
    \item Storing the information in a Program Knowledge Base (PKB), 
    \item Providing the user with the means to run queries against the PKB, written in an SQL-like Program Query Language (PQL), 
    \item Processing the PQL queries based on the information found in the PKB and providing the results to the user.
\end{itemize}

\begin{figure}
    \centering
    \includegraphics[width=0.9\linewidth]{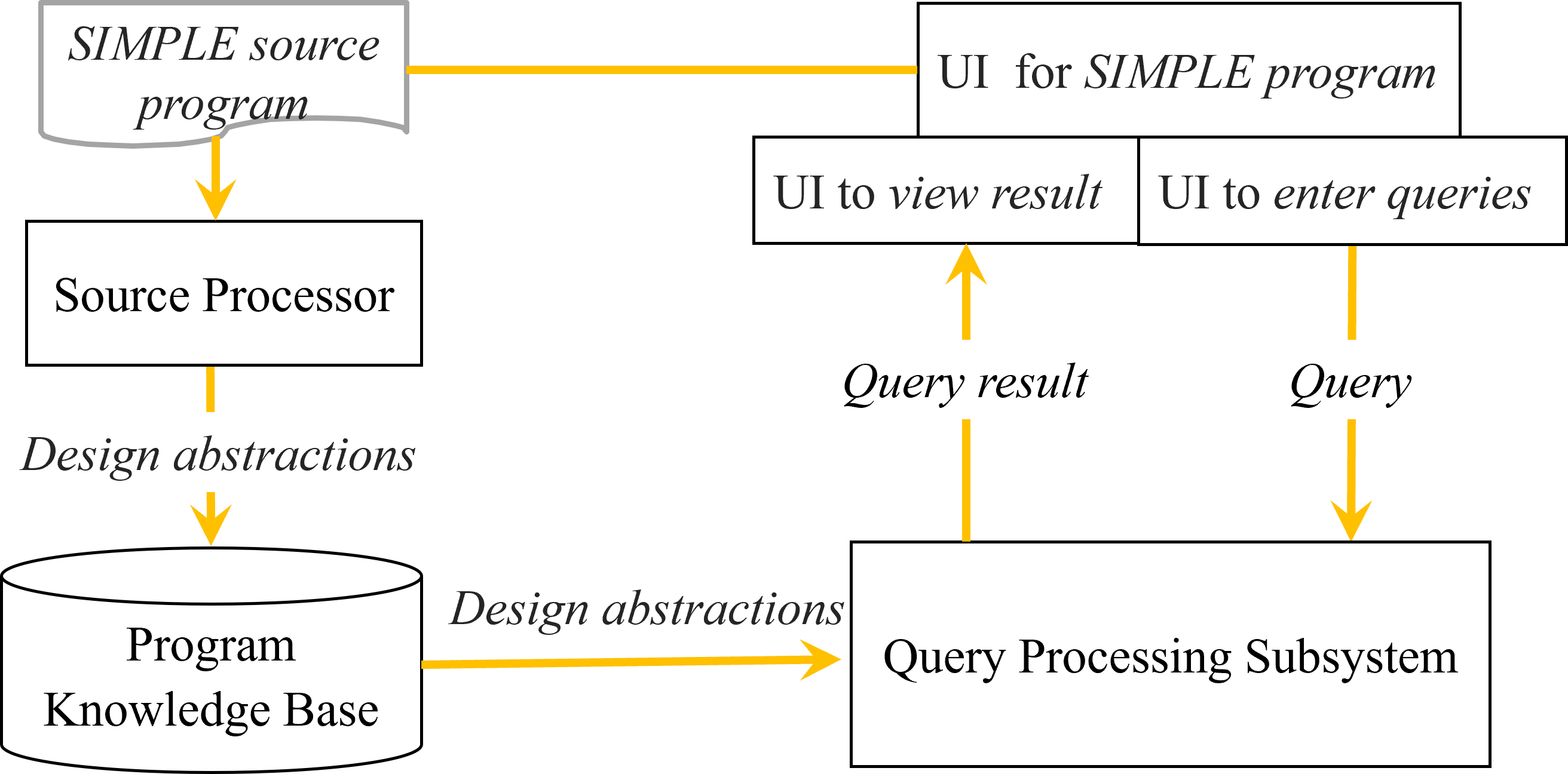}
    \caption{High-level software architecture of the Static Program Analyzer (SPA), the software to be developed by the students}
    \label{fig:spa-arch}
\end{figure}

Figure \ref{fig:spa-arch} depicts the high-level overview of the static program analyzer the students are expected to implement, with project requirements released in two parts. Throughout the project students engage in sprint planning, stand-up meetings, sprint reviews, and sprint retrospectives, mimicking the agile development methodology. They employ industry-standard tools like JIRA\footnote{https://www.atlassian.com/software/jira} for project management, and conduct stand-up meetings twice a week, following industry practices that ensure a realistic learning experience. The project timeline includes three major milestones distributed across five sprints, with an additional preparatory `Sprint Zero' at the beginning and a final `Release Sprint' before the final project delivery.

\begin{figure*}[t]
    \centering
    \includegraphics[width=1\linewidth]{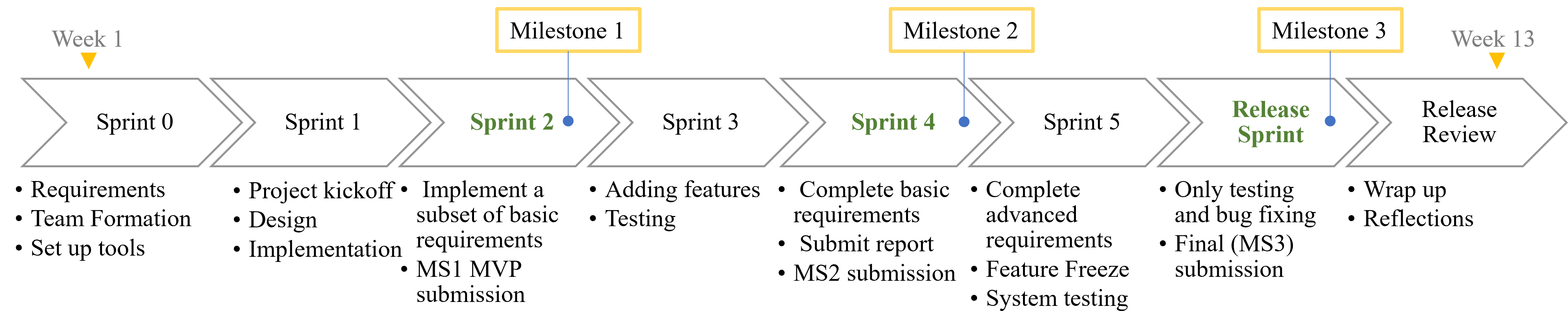}
    \caption{The timeline of the software engineering project, highlighting important milestones along with specific tasks in each phase. The development activities for the project span from Week 1-13, with three key milestones that need to be achieved by the students.}
    \label{fig:project-timeline}
\end{figure*}

The entire project timeline is illustrated in Figure \ref{fig:project-timeline}.  In Sprint Zero, project teams are formed, and students will take this time to understand the requirements, the team's strengths and weaknesses, etc. From Sprint One onwards, they work on the project requirements, and at the end of Sprint Two, they make the first project delivery (i.e. Milestone 1). Subsequent submissions happen at the end of Sprint Four and the final Release Sprint. 

The workload for the project is around 20 hours per team member per week. This accumulates to 240 person-hours per team per two-week sprint. Major evaluation criteria for the project include functional correctness, memory and computation efficiency, code quality (through the use of software engineering (SE) principles and design patterns), effective communication (oral and written communication), and continuous integration.

A fair amount of testing needs to be in place to achieve high functional correctness. Students must follow correct design principles and good coding practices, in order to produce a high-quality software product at the end of the project. Design principles such as the Single Responsibility Principle (SRP), and Do not Repeat Yourself (DRY), are some of the key principles among others that students are encouraged to follow. Multi-level code reviews are also in place to evaluate students' coding practices. 

The emergence of LLMs has increased the popularity of generative models for coding tasks, such as code infilling, bug fixing, and program comprehension among many others. Thus, we have encouraged students to use such generative models to potentially improve their productivity and efficiency. To analyze the usage of LLMs (i.e. ChatGPT, Copilot), we instructed students to annotate such generated code and document the prompts used. Section~\ref{sec:ai-gen-code} explains the annotation and documentation process, along with how we used the data for analysis. In our study, unlike most undergraduate courses where the usage of LLMs is explicitly prohibited, we actively allowed students to employ LLMs if they believed these tools would enhance their productivity. We did not apply any additional positive reinforcement to influence students' decisions.

\subsection{AI-Generated Code}\label{sec:ai-gen-code}
The students in this course were given the prerogative to use LLMs to generate code that can be integrated into their software development. However, as part of their code submissions, students were instructed to annotate such code with information describing the generative model. The annotations contained information such as the type of generator used (copilot, chatGPT-3.5, and chatGPT-4), as well as the level of \textit{human intervention} (0, 1, and 2) performed on the code snippets after receiving them from the generator. 

We define \textit{human intervention} as the effort required by the student to integrate AI-generated code into their software project. We measure this metric by quantifying the efforts and labeling them at one of three different levels:
\begin{itemize}
    \item \textbf{Level 0 (no intervention):} the AI-generated code is directly used in the project with \textbf{zero} changes by the student.
    \item \textbf{Level 1 (minor intervention):} the AI-generated code is used in the project, with $\leq 10\%$ of the lines being changed by the student.
    \item \textbf{Level 2 (major intervention):} the AI-generated code is used in the project, with > 10\% of the lines being changed by the student.
\end{itemize}


Listing 1 illustrates an example of annotating the AI-generated code. In this example, lines 1 to 4 are student-written, with the rest generated using chatGPT-3.5 (tagged as gpt3) and used with zero changes by the student. The annotations \verb+ai-gen start+ and \verb+ai-gen end+ were treated as reserved words and should not appear anywhere else in the entire SPA project, and the parentheses following \verb+ai-gen start+ contain information on the generator used, and the level of human intervention. 
We automatically collected relevant data by extracting from the code submissions at the three key milestones in the project timeline.

\begin{lstlisting}[caption={An example illustrating how comments are used to annotate AI-generated code.},captionpos=t, xleftmargin=3.5ex]
#pragma once
#include <utility>
#include <vector>
#include <string>
// ai-gen start (gpt3, 0)
struct vectorHash {
  std::size_t operator () (const std::vector<std::string> vect) const {
    std::size_t seed = 0;
      for (const std::string& str : vect) {
        seed ^= std::hash<std::string>{}(str)+0x9e3779b9 + (seed << 6) + (seed >> 2);
      }
    return seed;
  }
};
// ai-gen end

\end{lstlisting}

\subsection{User Study}
\begin{table*}[t]
\centering

\begin{tabular}{l l}
\hline
Total&139 respondents from 37 different groups.\\ 
Workload & 102 (74.1\%) were within the recommended workload for a semester, while 36 (25.9\%) exceeded the recommended workload.\\
Internships& 12 (8.6\%) had none while 84 (60.4\%) had 1, 33 (23.7\%) had two, 8 (5.8\%) had three, and 2 (1.4\%) had four internships.\\
Paid chatGPT& 15 (10.8\%) had a paid subscription to chatGPT, 124 (89.2\%) were using the free version. \\
\hline
\end{tabular}
\caption{Demographics of survey respondents}
\label{tab:demog}

\vspace*{-1\baselineskip}
\end{table*}

\begin{figure}[h]
\centering
\includegraphics[width=0.45\textwidth]{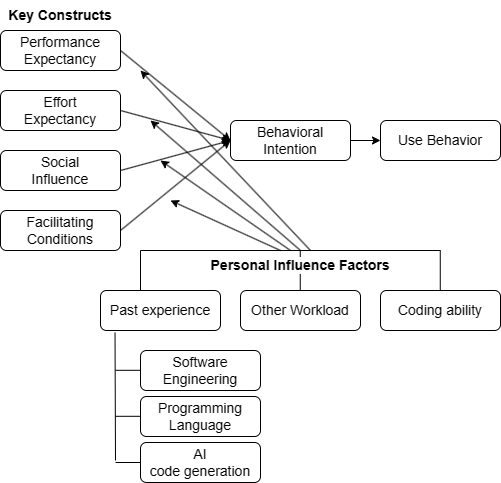}
\caption{Unified Theory of Acceptance and Use of Technology (UTAUT) model structure for the perception study} \label{fig:UTAUT}
\end{figure}

We conducted an online survey using the Unified Theory of Acceptance and Use of Technology (UTAUT) model by Venkatesh et al. \cite{V2003}. The study aims to capture the interaction between behavioral intention and the usage behavior of AI code generation, incorporating key constructs such as Performance Expectancy (PE), Effort Expectancy (EE), Social Influence (SI), and Facilitating Conditions (FC). The model is further enriched by incorporating moderating factors called personal influence factors. Namely past internship experience, student workload, and coding ability. An illustration of the complete model can be found in Figure \ref{fig:UTAUT}.

The survey included two open-ended questions to gather students' general perceptions on how the popularity of LLMs affected the student as a computer science undergraduate and as a prospective software engineering practitioner. These open-ended questions will be analyzed using the VADER (Valence Aware Dictionary, and sEntiment Reasoner) \cite{hutto2014vader} model for sentiment analysis.

\begin{figure}[b]
    \centering
    \includegraphics[width=\linewidth]{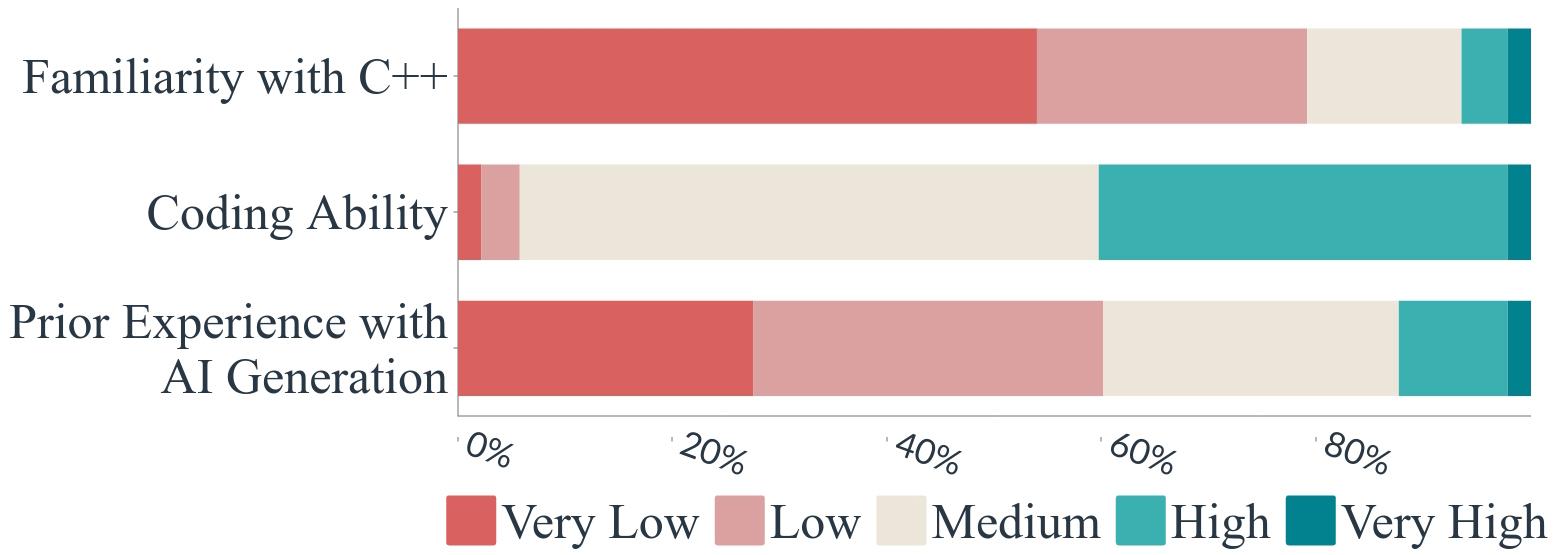}
    \caption{Self Assessed Personal Traits}
    \label{fig:bar}
\end{figure}

The survey was conducted in November 2023, after students completed their semester-long development effort. The online survey invited voluntary participation resulting in 139 respondents. Demographic details of the respondents are presented in Table \ref{tab:demog}. Familiarity with the programming language (C++), self-assessed coding ability, and prior experience with generating code using AI tools was assessed using the 5-point Likert scale questions in the survey. The results for these are depicted in the Figure \ref{fig:bar}.

To ascertain survey consistency, we carried out a reliability study for the key constructs in the model. The results are shown in Table \ref{tab:reliability}, indicate robust reliability, with all Cronbach's Alpha values exceeding 0.7, suggesting the survey questions are reliable.

\begin{table} [h]
\centering
\begin{tabular}{|l |l|} \hline  
UTAUT Key Construct & Cronbach's Alpha \\ \hline 
Performance Expectancy & 0.811\\ 
Effort Expectancy & 0.770\\ 
Social Influence & 0.739\\ 
Facilitating Conditions & 0.868\\ \hline

\end{tabular}
\caption{Reliability Analysis of Survey (n=139)}
\label{tab:reliability}

\vspace*{-1\baselineskip}
\end{table}

The analysis will delve into the application of the UTAUT model to analyze the use of LLMs for code generation. By leveraging the key constructs along with the moderating factors, this study seeks to understand the dynamics influencing behavioral intention and usage behavior in the context of AI-assisted code generation.

\section{Results}\label{secres}

\subsection{AI code generation}



214 students were enrolled in the course, which led to 37 teams of roughly six students in each team. Out of the 37 teams, 15 teams (40.5\% of total teams) indicated using AI-generated code in their project. Seven out of these 15 teams, or 19\% of all teams, accounted for 75\% of final AI-generated code snippets and 85\% of final AI-generated lines present in the final submission. 

Three AI code generators were used by the students: GitHub Copilot, ChatGPT-3.5 (free version) and ChatGPT-4 (paid subscription). The majority of the generated code snippets were from ChatGPT-3.5, which amounted to 82.5\% of the total. Whereas 9.5\% of generated snippets were from GitHub Copilot and 8\% were from ChatGPT-4.


The total lines of code generated were tracked across the three main deliverables of the project, Milestone 1 (MS1), Milestone 2 (MS2), and Milestone 3 (MS3). Table \ref{tab:milestone_code} depicts the number of code snippets and the average lines per code snippet in each of the milestones. 

The most number of AI-generated code snippets is observed in MS1 (53), whereas additional code snippets introduced in MS2 was 3 and for MS3 was 7. This indicates that students found AI code generators more useful in getting initial code structures, such as basic algorithms and design patterns generated. 

Another trend we observed was the average number of lines per generated code snippet kept increasing from MS1 to MS3.\footnote{One-liners generated by LLMs are less relevant here, as the average output is typically around 30 lines} This could be due to students gaining experience in how to better prompt the AI generator to get larger code snippets generated. This observation is also corroborated in our user perception study at the end of the semester.

\begin{table} [h]
\centering
\small
\begin{tabular}{|l |l|l|l|} \hline  
Milestone & Snippets & Lines & Lines / Snippet \\ \hline 
MS1 & 53 & 1781 & 33.60 \\ 
MS2 & 3 & 159 & 53.00 \\ 
MS3 & 7 & 513 & 73.29 \\ \hline

MS1 (aggregate) & 53 & 1781 & 33.60 \\ 
MS2 (aggregate) & 56 & 1940 & 34.64 \\ 
MS3 (aggregate) & 63 & 2453 & 38.94 \\ \hline

\end{tabular}
\caption{Statistics of AI-generated code (sum of all teams) over three milestones.}
\label{tab:milestone_code}

\vspace*{-1.5\baselineskip}
\end{table}



\begin{figure}[h]
    \centering
    \includegraphics[width=0.85\linewidth]{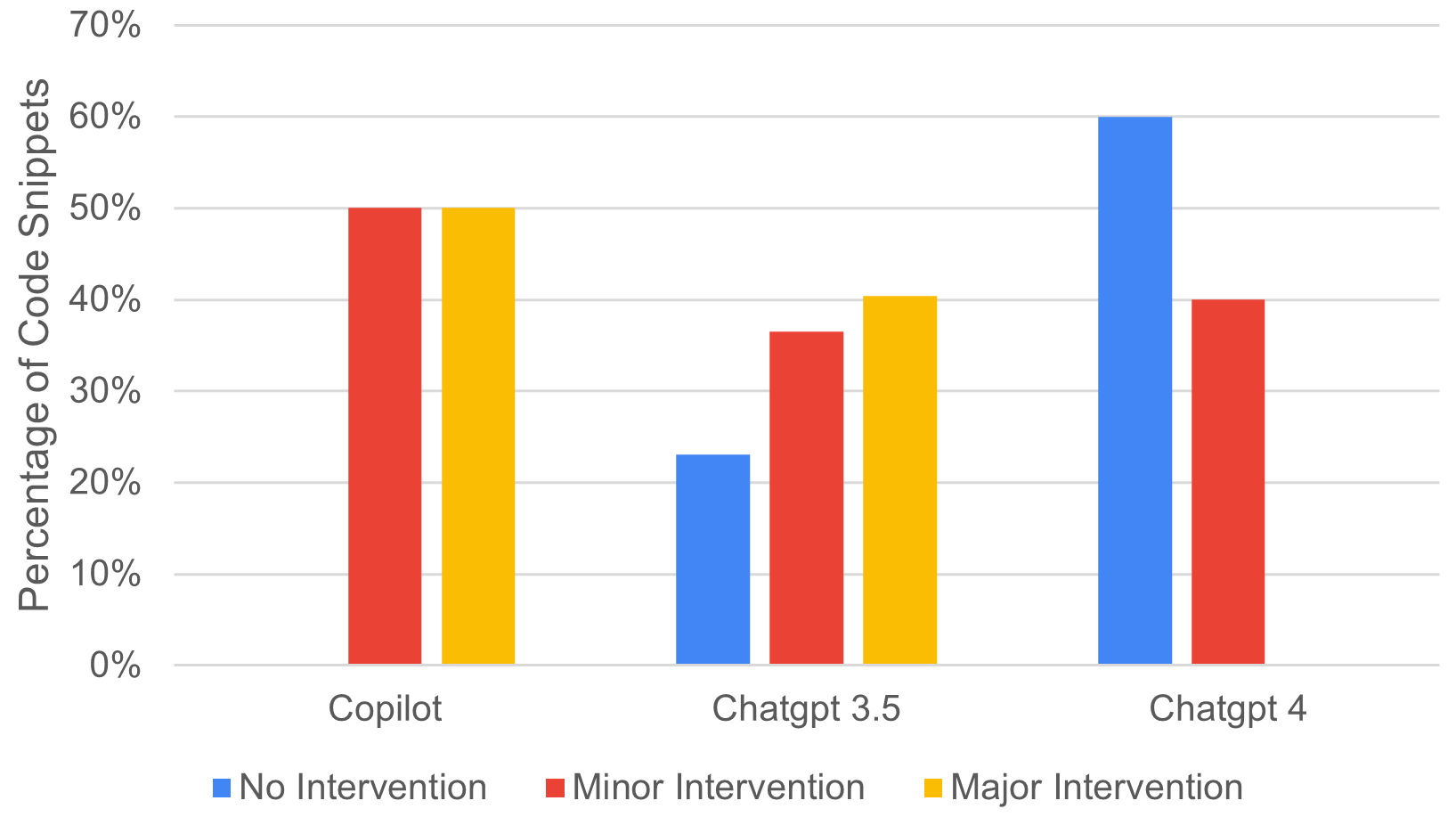}
    \caption{Levels of intervention performed on different generators' output}
    \label{fig:intervention_pctg}
    \vspace*{-1\baselineskip}
\end{figure}

We further analyze the effort needed to integrate AI-generated code in a practical software engineering project. We measure this using the \textit{human intervention} levels required for different generated code snippets, which were analyzed based on the annotations.
Figure \ref{fig:intervention_pctg} highlights the different levels of \textit{human intervention} the students performed after receiving the AI-generated code snippet. We can observe that all code snippets generated by Copilot needed extra \textit{human intervention}. It's also important to note that, the usage of Copilot tends to lead to the generation of much more simpler code snippets, instead of complex code snippets. In stark contrast, the code generated by the paid version of ChatGPT (v4) did not require students to perform major changes before using the generated code snippet in their submission, whereas the free version of ChatGPT saw intervention at all three levels. It is also noteworthy that the free version of ChatGPT was the most frequently used generator in this study.

\subsection{Prompt engineering}

\begin{table*}
\centering
\small
\begin{tabular}{l l l p{5.5in}}  \hline
& Prompts & (\%)& Example Prompts \\ \hline
(1) & 5 & 8.19& 1.1 // student code removed  \newline Is there a way to avoid repetition among the 3 different methods processStmtList, processWhileStmtList and processIfStmtList? \newline 1.2 // student code removed \newline Refactor the above code to use DFS to check for cyclic calls \\ \hline
(2) & 14 & 22.95& 2.1 make a regex string that matches all of these and nothing else >, <, >=, <=, ==, != or any whitespace \newline  2.2 Cpp code to print nested vectors \\ \hline
(3) & 11 & 18.03& 3.1 unordered\_set<unordered\_map<string, string>> can you help me implement the hash and equal function for this unordered set \newline  3.2 Cpp code to get all keys of unordered\_map \\ \hline
(4) & 24 & 39.34& 4.1 I have a arithmetic expression like "var1 + var2 * var3 + var4". Write a function to turn this into a list of prefix expressions ["+", "+", "var1", "*", "var2", "var3", "var4"] \newline  4.2 Write a DFS to detect cycles given an unordered\_map of nodes to its unordered\_set of connected nodes. \\ \hline
(5) & 7 & 11.47& 5.1 Given “<excerp>”, why do I get “<error>” \newline  5.2 write unit test for pattern manager class with catch2 \\ \hline

\end{tabular}
    \caption{Prompt distribution across categories and some example prompts from each category. (Note: we chose to show example prompts with a smaller number of words here for brevity)}
    \label{tab:promptdist}
    
\vspace*{-1.5\baselineskip}
\end{table*}
A total of 61 prompts used by the students were submitted, and these were recorded in a separate submission apart from the main code. The prompts used by the students were not captured exhaustively, as this was an optional requirement done voluntarily. We analyze the available prompts to understand the prevalent trends in prompting LLMs for code-generation tasks. All the prompts were pre-processed by removing code segments within the prompt, leaving only the natural language description to explain the scenario. 

\begin{figure}[h]
    \centering
    \includegraphics[width=0.60\linewidth]{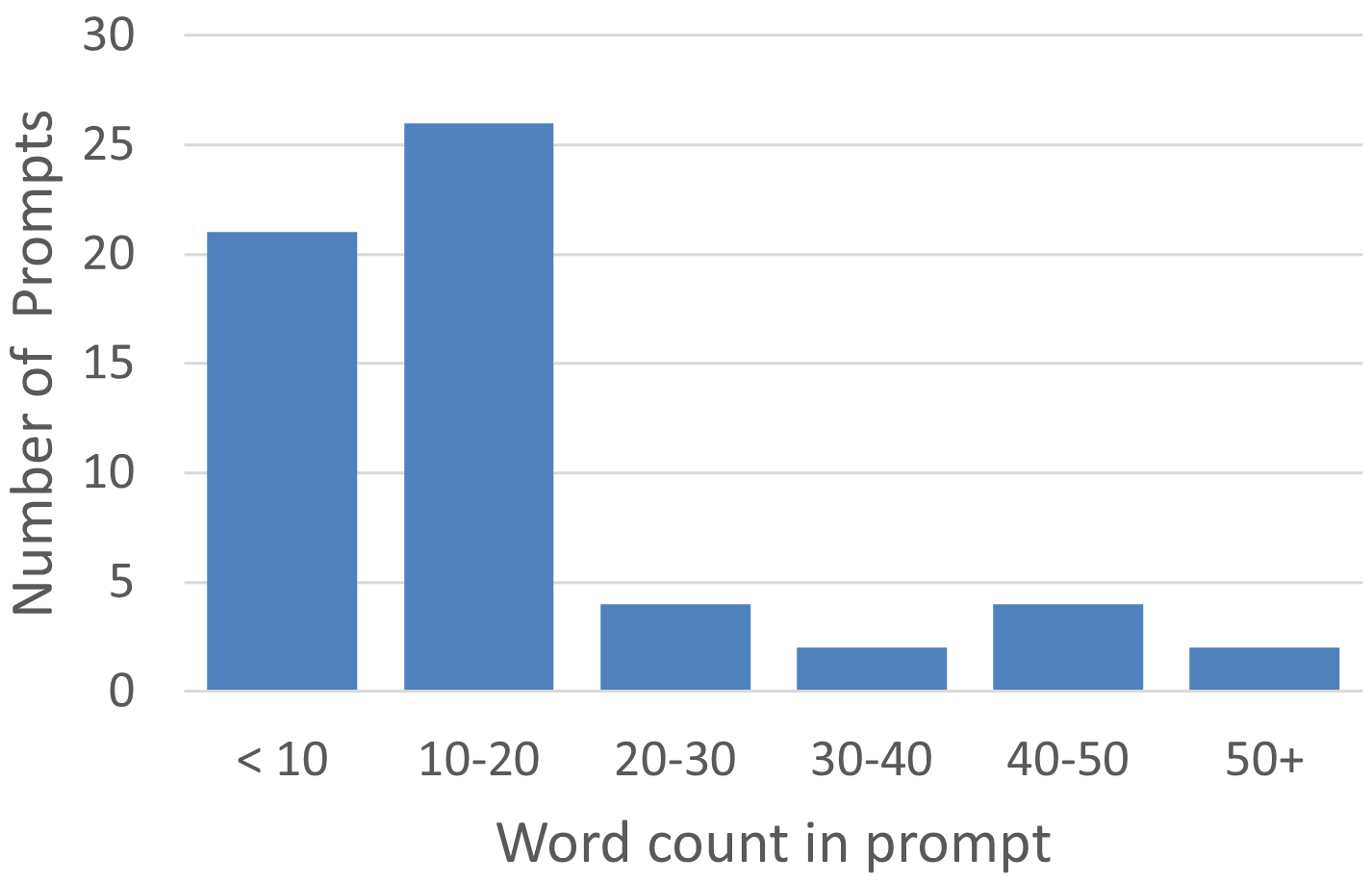}
    \caption{Word counts in prompts used for code generation}
    \label{fig:word}
    \vspace*{-1\baselineskip}
\end{figure}

The average word count of prompts used by the students stood at 18.0 words per prompt ($\sigma=17.6$). The breakdown of word counts in the prompts for code generation is shown in Figure \ref{fig:word}. The smallest prompt was 5 words, while the largest prompt was 103 words long. This shows the wide range of prompts used to give the context and requirements to the LLM for code generation. Analyzing the prompts qualitatively, we have categorized the prompts broadly into 5 categories. 

\begin{enumerate}
    \item  Improve student-written code (correctness/code quality)
    \item Simple tasks and help in C++
    \item Data structures for a specific requirement
    \item Solutions/algorithms for common problems
    \item Others
\end{enumerate}

Category (5) includes prompts to generate test cases, and design patterns as well as prompts to fill knowledge gaps. Table \ref{tab:promptdist} shows the breakdown of prompts in each category along with some examples. We can see the most usage of AI code generation has been for simple tasks in C++ along with generating solutions/algorithms for specific problems.

\subsection{Perceptions}

The behavioral intention and usage were measured using the online survey. On a 5-point Likert scale, (0 - strongly disagree, 5 - strongly agree) behavioral intention had $\mu=3.8$ ($\sigma=0.9$) and usage behavior had $\mu=3.2$ ($\sigma=1.2$).
These results indicated a high behavioral intention and usage behavior. However, the high variance also indicates there are other factors affecting this decision. We study other factors affecting the AI code generator usage, through the key constructs and personal influence factors.

\begin{table}[b]
\centering
\small
\caption{Correlation Analysis of Key Constructs (n=139, $\alpha=0.05$). $p < 0.001$ for all cells.}
\begin{tabular}{| l | l l l l l l |}   \hline
 & Usage& Intention & PE & EE & SI & FC \\ \hline
Usage& 1.00 &  &  &  &  &  \\
Intention & 0.60 & 1.00 &  &  &  &  \\
PE & 0.66 & 0.65 & 1.00 &  &  &  \\
EE & 0.65 & 0.59 & 0.87 & 1.00 &  &  \\
SI & 0.41 & 0.63 & 0.76 & 0.76 & 1.00 &  \\
FC & 0.73 & 0.68 & 0.72 & 0.63 & 0.57 & 1.00 \\
\hline
\end{tabular}

\label{tab:correlation}

\end{table} 

\begin{table*}
\small
    \centering
    \begin{tabular}{lp{6in}l}
\hline
          H4*&Students who rated themselves with lower coding skills show lower AI code usage ($\mu=3.14$ compared to $\mu=3.41$)& P=0.09\\
          H5**&Students with prior experience using AI code generators show a much higher behavioral intention and usage of AI. ($\mu=2.91$ compared to $\mu=3.76$)& P=0.0000\\
          H6**&Students who gained experience using AI code generation in this project show much higher intent to use it in future software engineering projects. ($\mu=2.8$ compared to $\mu=3.81$)& P=0.0000\\
 H7**& Students who have access to a paid subscription to chatGPT show a much higher behavioral intention and usage of AI. ($\mu=3.14$ compared to $\mu=4.13$)&P=0.0014\\\hline
    \end{tabular}
    \caption{Hypothesis testing.  *  statistical significance at $\alpha=0.1$, ** statistical significance at $\alpha=0.05$}
    \label{tab:hyptest}

    \vspace*{-2\baselineskip}
\end{table*}
Table \ref{tab:correlation} highlights strong correlations among the key constructs, with statistical significance for the UTAUT model. There is a high correlation between behavioral intention and usage behavior. Among the key factors, the highest correlation is with Facilitating Conditions (FC), Performance Expectancy (PE), and Effort Expectancy (EE). This shows that the AI code generation usage depends on having access to these services (FC), a reduction in the effort required to write high-quality software by the use of code generation (PE), and the ease of prompt engineering to generate required code (EE). The correlation is lowest with Social Influence (SI), which indicates that the intent or usage of AI code generation is less influenced by social factors such as what one's peers would think about their usage of code generators. The following null hypotheses were tested about the personal influence factors on the behavioral intention and usage (BI/U) of AI code generators.

\begin{itemize}
    \item \textbf{H1}: There is no difference in BI/U between the students who had the recommended workload and those who exceeded it. 
    \item \textbf{H2}: There is no difference in BI/U between the students who have higher internship experience ($> 1$) and those who have less internship experience.
    \item \textbf{H3}: There is no difference in BI/U between the students who rated themselves with higher C++ familiarity ($>1$) and those who rated themselves with lower C++ familiarity.
    \item \textbf{H4}: There is no difference in BI/U between the students who rated themselves with lower coding skills ($<=3$) and those who rated themselves with higher coding skills.
    \item \textbf{H5}: There is no difference in BI/U between students who had no prior exposure using AI code generators and students with prior exposure using AI code generators
    \item \textbf{H6}: There is no difference in BI/U between students who used AI in this project compared to students who did not use AI in this project.
    \item \textbf{H7}: There is no difference in BI/U between students with a paid subscription to chatGPT and students without a paid subscription to chatGPT.
\end{itemize}

The t-tests were carried out between the different populations for H1 - H7, and it is shown that there were no statistically significant differences for H1, H2, and H3. However, H4 - H7 all had statistically significant differences between populations. These results are summarized in Table \ref{tab:hyptest}. 
H4 shows that if a student rated themselves to have less coding skills they were less willing to use AI code generation. This might seem counter intuitive at a glance, due to an expectation for less-skilled students to seek out tools to aid them in their work. However, this could be due to the students being less confident in their ability to read and understand code generated by AI for correctness and code quality. Which encourages them to write the code by themselves rather than rely on generated code. This also exposes the need for future SE practitioners to be highly skilled in coding to be able to make use of these new technologies effectively in their work.

H5 and H6 both highlight that the usage of AI code generators is also a skill that needs to be learned and practiced. By gathering experience in using AI generators, students are allowed to experiment and learn about proper prompt-engineering techniques, as well as how to do manual interventions to the generated code block before using it in their project. Exposing computer science students to these tools allows them to grow and learn when and where to apply these tools in large-scale SE projects in the future.

Finally, H7 highlights the tendency of paid ChatGPT users to use it more than students who use the free version. This could be due to the superior performance of the paid version. However, this could also be a biased observation as the paying customers are those who have more exposure to and experience using code generation tools, which can be a higher influencing factor.

\textbf{Sentiment Analysis:}
We further asked students additional questions to understand their perception of generative models that can generate code. The first question inquires about the perception as an undergraduate student, \textit{Q1 - How does AI code generation impact you as a Computer Science student?} 120 students responded to this question. VADER analysis shows that 90 (75\%) of the responses were positive, while 13 (10.8\%) were negative and 7 (5.8\%) were neutral. 
Some positive use cases highlighted by the students are; good for understanding basic framework concepts and language syntax, simplifying repetitive tasks, debugging and understanding errors in code, and making coding/learning more efficient. 
Some negatives highlighted by the students are;  taking away the enjoyment of coding, good for scripting, bad for large projects and heavy reliance on AI Code generation will diminish the critical and logical thinking skills that CS students should acquire.

The second question captures their sentiment as future practitioners of software engineering,
\textit{Q2 - How would AI code generation impact you as a Software Engineer in the future?} 109 students responded to this question. VADER analysis shows that 71 (65\%) of the responses were positive, while 26 (23.8\%) were neutral and 12 (11\%) were negative. We observe that the future Software Engineering industry outlook has relatively high negative sentiments compared to the previous question. 

Some key positives listed by the students include; how AI code generation will make them more efficient Software Engineers, allowing them to tackle more work. Most positive commenters saw themselves utilizing AI as part of their toolset as software engineers. The most common concern raised by the negative sentiments was the impact on the job market leading to fewer programming jobs. Most concerned respondents highlighted the need to upskill and differentiate themselves further to survive in the Software industry in the future.

\subsection{Correctness and Quality}

Since the project is evaluated in an academic setting, each project team's code base is evaluated for its correctness using test cases, and its performance using benchmarking, with industry-standard code quality checks also being employed. Utilizing these evaluations, we analyzed if there was any difference between teams who utilized LLMs for code generation and teams who did not. 

The analysis showed that there is no statistically significant difference between the two categories of teams. This indicates that diligent vetting and human intervention of AI-generated code, before usage in the student code base has proven to be effective. Therefore, AI-generated code is capable of achieving a similar level of accuracy, performance, and code quality with proper human intervention.

\section{Discussion}
There was a clear reduction in LLM usage as the project progressed. While the initial stages of the project in the first milestone saw the highest utilization of LLMs, subsequent milestones witnessed a decrease in utilization but an increase in complexity of the generated code. This trend suggests that LLMs helped provide a quick start to the project, especially in overcoming the steep learning curve. As students become more familiar with the programming language and project requirements, and as the complexity of the code base increases, the benefits of using LLMs diminish and result in fewer instances of usage. Interestingly, in the later stages, LLMs were typically tasked with generating more complex code compared to the simpler tasks they were used for at the initial stage of the project.

Analyzing the correlations between AI code generation usage and personal influence factors showed that students more skilled in coding were more inclined to use AI code generation tools, whereas students with less coding skills were more averse to it. This is counter intuitive to the initial idea that students less skilled in coding would be more dependent on AI tools to generate code. This could be due to the added skill requirement to read and understand the correctness and quality of AI-generated code, as well as having enough experience to perform the necessary human intervention to manipulate the generated code into a usable format. This finding highlights solid foundational coding skills remain a vital factor for software engineers looking to benefit from AI code generators.

We observe that a smaller proportion of the cohort was responsible for the majority of the AI-generated code. The observation aligns closely with the Pareto principle, where approximately 80\% of the generated code was contributed by 20\% of the teams. This disparity could be due to the varying levels of familiarity and proficiency with using AI generator tools for code generation among students. Comparing the correctness and quality of code bases among teams that heavily used AI tools, teams that used them lightly, and teams that did not use AI tools at all, the results revealed no significant difference in the quality and correctness of code between them. This finding suggests that, with proper human intervention, the use of AI tools does not inherently compromise the quality and correctness of code in software development projects. Linking this to the earlier observation of the Pareto principle, the disparity in AI tool usage did not translate to a distinct difference in the overall quality of the code, and this observation has profound implications for educators.

From an educator's perspective, it becomes crucial to empower students with the skills necessary for successful human-AI collaboration, this includes enhancing students' abilities in prompt engineering, interpreting and manipulating AI-generated code, to enable them to leverage these tools more effectively. The end goal is to ensure students can critically assess and integrate AI-generated code into their projects, instead of becoming passive bystanders. An increased focus on these skills is vital in preparing students who are adept at using new technology and tools to their full potential, without becoming overly dependent on them or compromising the quality of their work.


\section{Related Work}\label{sec-lit}




There are several general-purpose large language models (LLMs) that are available, such as Google's BARD and OpenAI's GPT-3 and GPT-4 \cite{10263706}. 
AI-powered code generators help streamline coding processes, automate routine tasks, and even predict and suggest code snippets. 
Developed by GitHub in collaboration with OpenAI, GitHub Copilot \cite{Copilot}  aids developers in writing better code at an expedited pace. Several papers (\cite{MORADIDAKHEL2023111734},  \cite{peng2023impact} and \cite{copilot1}) discuss Copilot's role and effect in code generation under different user environments. Amazon's CodeWhisperer \cite{yetiştiren2023evaluating} revolutionizes the coding process by offering real-time suggestions ranging from snippets to entire functions. 
There are many other works in the literature that provide a brief overview of recent advancements in AI code generators including \cite{hou2023large}, \cite{fan2023large}, \cite{10.1145/3491101.3519665},  \cite{10.1145/3558489.3559072} and \cite{yetiştiren2023evaluating}.  While \cite{wang2023software} analyzes 52 relevant studies that have used LLMs for software testing, from both the software testing and LLMs perspectives.  In \cite{kazemitabaar2023novices}, \cite{zheng2023understanding} and  \cite{10213396}, authors examine the use of LLM-based code generators in software engineering through literature review, programming tasks, or surveys.  The paper \cite{10.1145/3491101.3519665} conducted a user study with 24 participants to understand the role of Copilot in programming assignments. Several studies in the recent past (\cite{10.1145/3573381.3603362}, \cite{10298678} and \cite{Prather_2023}) have looked at the perception of LLM-based code generators in different settings, such as those of IT department employees, early adopters, and novice programmers. They are studied in a smaller user base.

Our work significantly differs from all these existing works in multiple aspects. We conducted a study of the effect of code generators on software engineering projects in a controlled setting of 214 users, differing from related works in terms of the scale of the study, patterns of usage, sentimental analysis, the role of user demographic details in generated code usage, and so on. 

\section{Summary}\label{sec-con}

\textbf{Research objectives and contributions:} 
Our research presents the first comprehensive study on the use of LLMs in academic software engineering projects, and examines the perceptions of students who have used LLMs for their projects. Our study reveals insights on how the usage of AI tools correlates with factors like coding skills and experience, and provides a trajectory for educators in software engineering to emphasize the effective usage of AI tools in education programs. 

\textbf{Summary of findings and implication:} 
Our study reveals that LLMs are significantly beneficial in the early stages of software development projects, enhancing productivity in routine tasks like debugging. However, the adoption of LLMs varied depending on the students' coding skills and prior experience with AI generators, indicating a learning curve in using these tools effectively. We also observed no significant difference in the correctness and quality of developed software between teams with heavy and no AI usage. \\
\textbf{Limitations:} Our study is subject to certain limitations, as not all students annotated their usage of AI-generated code, and not all prompts used for code generation were recorded, which could lead to potential skews in our results. Not all students responded to the survey, therefore our findings might not fully represent the population of the entire cohort's experience in using LLMs. The study was also conducted in an academic setting, and may not perfectly mirror constraints in the software industry, where data privacy and intellectual property concerns exist. Students with potential financial constraints may also have faced restricted access to paid versions of LLMs, potentially influencing the extent of their usage.

Despite these limitations, our research provides foundational insights into the role of LLMs in software engineering, shedding light on future exploration in both academic and industry settings.

\begin{acks}
Teaching team of CS3203 including Dr. Zhao Jin and Head TA Ng Tzerbin.
\end{acks}

\bibliographystyle{ACM-Reference-Format}
\bibliography{software}

\end{document}